\documentclass[reprint,showpacs,amsmath,amssymb,aps,superscriptaddress,floatfix,longbibliography,nofootinbib]{revtex4-2}
\usepackage{amsthm}
\usepackage{amsmath, mathtools}
\usepackage[utf8]{inputenc}	
\usepackage[english]{babel}
\usepackage{amsmath}
\usepackage{graphicx}		
\usepackage{natbib}
\usepackage{textcomp}
\usepackage{gensymb}
\usepackage[usenames,dvipsnames,svgnames,table]{xcolor}
\usepackage[hidelinks,colorlinks=false,urlcolor=Cerulean,citecolor=black]{hyperref}
\usepackage{siunitx}
\usepackage{mathrsfs}
\usepackage{multirow}
\usepackage{bm}
\usepackage{xfrac}
\usepackage{comment}
\usepackage[normalem]{ulem}

\renewcommand{\i}{\mathrm{i}}

\DeclareMathOperator{\R}{\mathbb{R}}
\DeclareMathOperator{\C}{\mathbb{C}}

\begin{document}

\title{Exact expression for maximum Lyapunov exponent during transients \\ in computationally powerful dynamical networks}

\author{Arthur S. Powanwe}
\thanks{Co-first authors}
\affiliation{Department of Mathematics, Western University, London ON, Canada}
\affiliation{Western Institute for Neuroscience, Western University, London ON, Canada}
\affiliation{Fields Lab for Network Computation, Fields Institute, Toronto ON, Canada}

\author{Luisa H. B. Liboni}
\thanks{Co-first authors}
\affiliation{Fields Lab for Network Computation, Fields Institute, Toronto ON, Canada}
\affiliation{King's University College at Western University, London ON, Canada}

\author{Anif N. Shikder}
\affiliation{Department of Mathematics, Western University, London ON, Canada}
\affiliation{Western Institute for Neuroscience, Western University, London ON, Canada}
\affiliation{Fields Lab for Network Computation, Fields Institute, Toronto ON, Canada}

\author{Alexandra N. Busch}
\affiliation{Department of Mathematics, Western University, London ON, Canada}
\affiliation{Western Institute for Neuroscience, Western University, London ON, Canada}
\affiliation{Fields Lab for Network Computation, Fields Institute, Toronto ON, Canada}

\author{Kalel L. Rossi}
\affiliation{Cellular Computations and Learning, Max Planck Institute for Neurobiology of Behavior, Bonn, Germany}

\author{Todd Coleman}
\affiliation{Department of Bioengineering, Stanford University, Stanford CA, USA}

\author{Ján Mináč}
\affiliation{Department of Mathematics, Western University, London ON, Canada}
\affiliation{Fields Lab for Network Computation, Fields Institute, Toronto ON, Canada}

\author{Ulrike Feudel}
\affiliation{Theoretical Physics/Complex Systems, ICBM, Carl von Ossietzky University Oldenburg, Oldenburg, Germany}

\author{Roberto C. Budzinski}
\thanks{Co-senior authors}
\affiliation{Department of Neuroscience, University of Lethbridge, Lethbridge AB, Canada}
\affiliation{Fields Lab for Network Computation, Fields Institute, Toronto ON, Canada}

\author{Lyle E. Muller}
\thanks{Co-senior authors}
\affiliation{Department of Mathematics, Western University, London ON, Canada}
\affiliation{Western Institute for Neuroscience, Western University, London ON, Canada}
\affiliation{Fields Lab for Network Computation, Fields Institute, Toronto ON, Canada}

\begin{abstract}
We study a network whose rich spatiotemporal dynamics have recently been shown to enable dynamics-based computation, including logic gates, short-term memory, and simple encryption. The network's time dynamics can be exactly solved through a nonlinear coordinate transformation. Here, we derive an exact analytical expression for the network's time-dependent maximum Lyapunov exponent (MLE). We demonstrate, both numerically and analytically, that the network exhibits positive MLEs during the transients that are useful for computation. Our framework enables algebraic manipulation of transient lifetimes through network connectivity and initial conditions, providing a rigorous theoretical foundation for understanding and controlling computation with transients.
\end{abstract} 

\maketitle

Transient dynamics are key to understanding computation in recurrent neural networks, where dense internal connections cause inputs to reverberate, building up rich activity states. Recent work has examined the internal dynamics of neural networks as they perform sequence processing \cite{lu2025state} or language tasks \cite{keller2025nu}. However, a general understanding of the dynamical regimes that are important for computing with transients remains lacking.

Here, we study a specific complex-valued network that bears similarity to networks of nonlinear Kuramoto oscillators \cite{muller2021algebraic,budzinski2022geometry,budzinski2023analytical}. This network can perform a range of computations, ranging from short-term memory and message encryption \cite{budzinski2024exact} to image segmentation \cite{liboni2025image} and short-term predictions of naturalistic movies \cite{benigno2023waves}. We have also recently shown that this network has a formal mathematical connection to the structured state-space sequence model (S4) \cite{cv5}, a leading machine learning architecture for sequence processing \cite{gupta2022diagonal, gu2022parameterization} and language generation \cite{gu2021efficiently,smith2022simplified}. Interestingly, it is possible to obtain an exact solution for the network's time dynamics \cite{muller2021algebraic}. This presents a unique opportunity: we can rigorously analyze the internal dynamics of this neural network, which is known to have not only substantial computational power, but also clear practical applications. This combination of mathematical tractability and computational power thus represents a key point for theoretical study in both nonlinear dynamics and computation in neural networks.

Here, we provide theoretical insight into this system. The network produces a rich diversity of spatiotemporal dynamics, including activity patterns that exhibit characteristics of spatiotemporal chaos. We investigate the possibility that spatiotemporal chaos underlies the computational power of this system by deriving an exact, closed-form expression for the maximum Lyapunov exponent (MLE) for individual dynamical trajectories -- the first time this has been achieved for a large nonlinear network. We demonstrate that there exist conditions under which the time-dependent MLE can be positive or negative, allowing us to understand when the network uses trajectories of each type for computation.

Taken together, these results provide a comprehensive theoretical basis for understanding why the network can perform computation in its transient dynamics. Spatiotemporal dynamics have been proposed to be a powerful substrate for computation in different systems, from engineered physical systems \cite{fernando2003pattern, adamatzky2002experimental, del2018leveraging} to biological neural networks \cite{ermentrout2001traveling, muller2018cortical} and artificial neural networks \cite{ricci2021kuranet,keller2023neural,miyato2024artificial,jacobs2025traveling,csaba2020coupled,todri2024computing}. Further, for the state-space models (SSMs) such as S4 that have recently been introduced for generative language applications, complex-valued SSMs specifically have expanded computational capacity over real-valued SSMs \cite{ran2024provable}. The theoretical framework we introduce here has the potential to provide fundamental insights into the computations of these systems that are central to modern machine learning tasks.

The network is defined by the equations on $N$ nodes:
\begin{widetext}
\begin{equation}
    \dot{\psi}_{i}(t) = \omega + \epsilon\sum_{j = 1}^{N} a_{ij}\Big( \sin{ \big(\psi_{j}(t) - \psi_i(t) - \phi \big)} -\i \cos{ \big(\psi_{j}(t) - \psi_i(t) - \phi \big)} \Big),
\label{eq:nonlinear_eq}
\end{equation}
\end{widetext}
where $\psi_{i}(t) \in \C$ defines the state of node $i$ at time $t$, and $a_{ij}$ is an element of the adjacency matrix that represents the connection from node $j$ to $i$. The coupling strength between nodes is given by $\epsilon \in \R$, $\phi$ represents a homogeneous phase delay between nodes, and $\omega$ is the intrinsic frequency of oscillation. Finally, $\i$ represents the imaginary unit.

We can now show that Eq.~(\ref{eq:nonlinear_eq}) admits an exact, closed-form solution \cite{muller2021algebraic,budzinski2022geometry,budzinski2023analytical}. After applying standard identities and rearranging algebraically (see Supplement Sec.~I), we obtain:
\begin{equation}
    \i\dot{\psi_i} = \i\omega + \epsilon e^{-\i\psi_i} \sum_{j=1}^{N} a_{ij}e^{-\i\phi} e^{\i\psi_j} \,.
\end{equation}
This system of equations is nonlinear in $\psi_i$; however, by applying the nonlinear coordinate transformation $x_i = e^{\i\psi_i}$ and noticing that $\dot{x}_i = \i e^{\i\psi_i}\dot{\psi}_i$, we obtain an exact solution in the new variable $x_i$ (see Supplement Sec.~I), expressed in matrix form as:
\begin{equation} \label{eq:solution}
    \bm{x}(t) = e^{\i\omega t}e^{\bm{K}t}\bm{x}(0),
\end{equation}
where $\bm{x}(t) \in \C^N$ is the state vector of the network at time $t$, $e$ here is the matrix exponential, $\bm{K} = \epsilon e^{-\i\phi} \bm{A}$ is a composite matrix that contains information about the adjacency matrix $\bm{A}$, coupling strength $\epsilon$ and phase-lag $\phi$, and $\bm{x}(0) \in \C^N$ are the initial conditions. Lastly, using the inverse coordinate transform $\Phi : \mathbb{C}^N \to \mathbb{C}^N$ with $\Phi(\psi) = e^{\i\psi}$ and $\Phi^{-1}(x) = -\,\i\,\ln x$, we obtain the solution in terms of $\psi_i$:
\begin{equation}
    \bm{\psi}(t) = -\i \ln{\Big(e^{\i\omega t} e^{\bm{K}t} e^{\i\bm{\psi}(0)}\big)},
\end{equation}
which expresses the time evolution of Eq.~\eqref{eq:nonlinear_eq} in closed form. In the following, we will work with the system under the nonlinear coordinate transformation, deriving an expression for the time-dependent MLE in terms of the variable $\bm{x}(t)$.
\begin{figure*}[thb]
    \centering
    \includegraphics[width=0.925\linewidth]{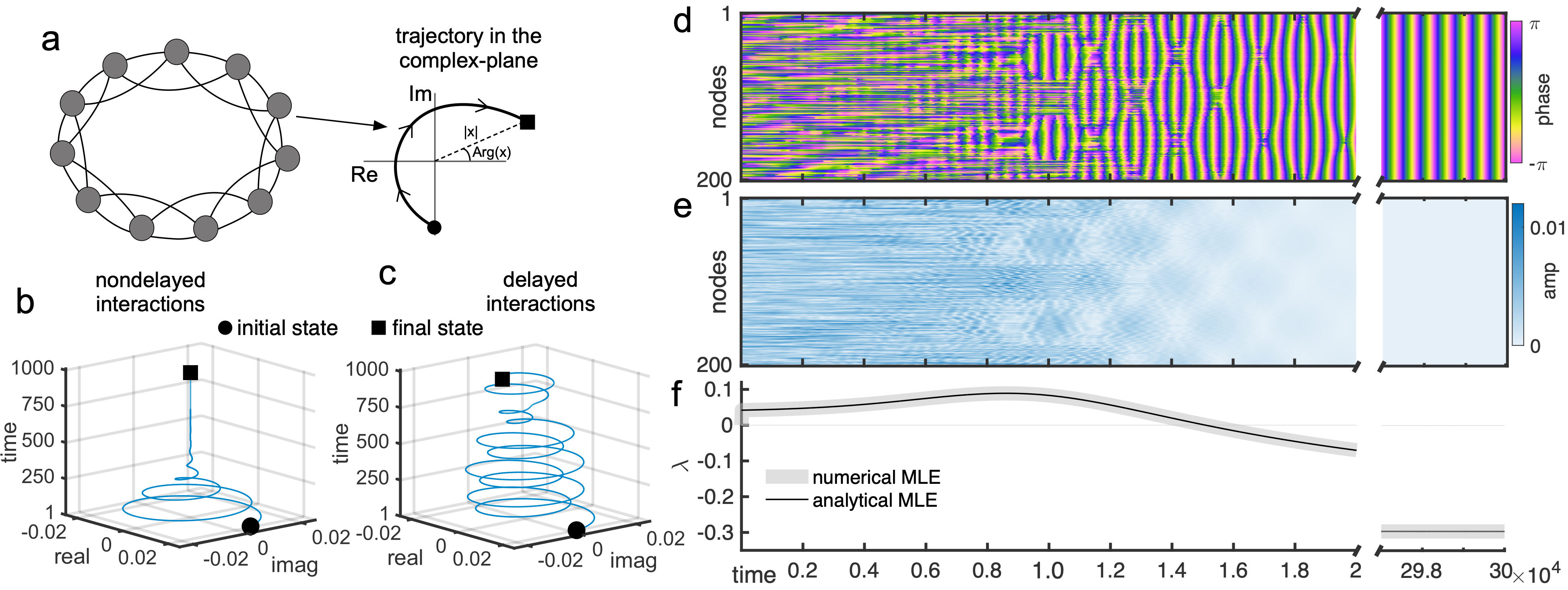}
    \caption{\textbf{The network displays rich spatiotemporal dynamics in its transient states.} \textbf{(a)} Nodes' dynamics is given by a complex-valued number, represented by phase (or argument) and amplitude under a hyperbolic coordinate frame. \textbf{(b)} In general, in a network with nondelayed interactions ($\phi = 0$), the dynamics will quickly converge to the origin (fixed point) - here we represented the behavior of a single node in the network. \textbf{(c)} For a network with phase-delayed interactions, the dynamics in much richer and the transient is longer. \textbf{(d), (e)} The network's spatiotemporal dynamics displays intricate patterns (both in phase and in amplitude) during transient. \textbf{(f)} We evaluate the time-dependent MLE, where we observe the MLE is positive during the transient, and it becomes negative as the network approaches the fixed point. Here, we consider a network on a $k$-ring graph with $N=201$ nodes, $k = 15$, $\omega = 0$, $\epsilon = 0.5$, and $\phi = 1.55$.}
    \label{fig:schematic_examples}
\end{figure*}

Under this coordinate transformation, the matrix exponential $e^{\bm{K}t}$ can now be seen to be an exact Koopman operator for this nonlinear networked system, working in the same space as the network dynamics $\C^N$ and acting on the state vector $\bm{x}(t)$ itself (so that the observable is the identity function \cite{brunton2021modern}). This provides the first reported example of a finite-dimensional Koopman operator that can be expressed in closed-form for a nonlinear networked system. The ability to represent the full nonlinear dynamics through this finite-dimensional operator enables exact analytical treatment of the system's temporal evolution without requiring infinite-dimensional function spaces typically associated with Koopman theory \cite{brunton2021modern}.

The temporal dynamics $\bm{x}(t)$ is then determined by the spectrum of $\bm{K}$: if any eigenvalue has a positive real part, the modulus of the node states $|x_i(t)|$ will, in general, diverge to infinity. These dynamics become much more tractable, however, in a hyperbolic coordinate frame. To do this, we utilize a M\"obius transformation $\bm{x}^{\mathrm{h}}(t) = 2\bm{x}(t)/(1 - \lVert \bm{x}(t) \rVert^2)$, where $\bm{x}^\mathrm{h}(t)$ is the dynamics in the new, hyperbolic coordinate frame, and $\lVert \bm{x}(t) \rVert$ is the Euclidean norm of $\bm{x}(t)$. Under this additional transformation, the closed-form solution Eq.~\eqref{eq:solution} takes the following form (see Supplement Sec.~II)
\begin{equation} \label{eq:solution_hyperbolic}
\bm{x}^{\mathrm{h}}(t) =\bigg(\dfrac{1 - \lVert \bm{x}(0) \rVert^2}{1 - \lVert \bm{x}(t)\rVert^2}\bigg) e^{\i\omega t}e^{\bm{K}t}\bm{x}^{\mathrm{h}}(0)\,.
\end{equation}
It becomes clear that,when $|x_{i}| \rightarrow \infty$ asymptotically in the Euclidean space, $|x_{i}^\mathrm{h}| \rightarrow 0$ in the hyperbolic space, bringing the fixed point of Eq.~(\ref{eq:solution}) from infinity to the origin of the complex plane.

We now have three different representations of the nonlinear dynamical system in Eq.~(\ref{eq:nonlinear_eq}) - $\bm{\psi}(t)$, $\bm{x}(t)$, and $\bm{x}^{\mathrm{h}}(t)$ - each with different utility for the theoretical analysis. We first performed a series of checks to validate the equivalence of these different representations. Specifically, numerical integration of the $\bm{\psi}(t)$ dynamics in Eq.~(\ref{eq:nonlinear_eq}) generates precisely the same dynamical trajectory as the dynamics of $\bm{x}(t)$ obtained through evaluation of the analytical expression Eq.~(\ref{eq:solution}) (Supplement Fig.~S1). Further, the dynamics in the Euclidean space $\bm{x}(t)$ is equivalent to the dynamics in the hyperbolic space $\bm{x}^{\mathrm{h}}(t)$ under the Möbius transformation (Supplement Fig.~S1). This confirms that these three representations of the system ($\bm{\psi}(t)$, $\bm{x}(t)$, and $\bm{x}^{\mathrm{h}}(t)$) are equivalent.

With this analytical approach established, we can now study the spatiotemporal dynamics generated by the network -- both in amplitude $|x_{i}^\mathrm{h}(t)|$ and phase $\mathrm{Arg}(x_{i}^\mathrm{h}(t))$ (Fig.~\ref{fig:schematic_examples}a). The phase delay $\phi$ becomes important here because it can enable the network to generate long transients. Specifically, a network with non-delayed interactions will, in general, quickly transition to the trivial fixed point where node state $|x_{i}^\mathrm{h}(t)| \rightarrow 0, \, \forall \,i$ in the hyperbolic frame (Fig.~\ref{fig:schematic_examples}b). With phase delays, however, transients can become rich and exhibit long timescales (Fig.~\ref{fig:schematic_examples}c). During this transient, the network can display intricate spatiotemporal activity patterns, with synchrony and asynchrony in phase and amplitude, resembling a classical chimera state (Figs.~\ref{fig:schematic_examples}d and \ref{fig:schematic_examples}e). As time evolves, the network transitions to the asymptotic state, with amplitude tending to zero. We note that, in the examples in Figs.~\ref{fig:schematic_examples}d and \ref{fig:schematic_examples}e, the amplitude is still finite, and the asymptotic state is only reached as $t \rightarrow \infty$.

We next sought to characterize the network dynamics during these long transients. To do this, we analytically compute the time-dependent MLE. The MLE is generally defined as the characteristic  separation rate of trajectories starting from infinitesimally close initial conditions. With this, the time-dependent MLE is defined as: $\lambda(t) = \lim_{\varepsilon \rightarrow 0} \frac{1}{t} \ln{ \Big( \frac{\lVert \delta \bm{x}^{\mathrm{h}}(t)) \rVert} { \lVert \delta \bm{x}^{\mathrm{h}}(0) \rVert} \Big)}$, where $\varepsilon$ is the infinitesimal perturbation over the initial condition, and $\delta \bm{x}^{\mathrm{h}}$ is the deviation between nearby trajectories in the hyperbolic space. Because we have a closed-form solution for the network dynamics in Eq.~\eqref{eq:solution_hyperbolic}, we are also able to obtain a closed-form solution for the time-dependent MLE (see Supplement Sec.~III for details):
\begin{equation}
      \lambda(t) =  \frac{1}{t} \ln{\left( \frac{\big(1 - \lVert \bm{x}(0) \lVert^2 \big)\sqrt{\sum\limits_{j=1}^N e^{2 \mathrm{Re}(\gamma_{j}) t} ~|\nu_j|^2}}{1 - \sum\limits_{j=1}^N e^{2 \mathrm{Re}(\gamma_{j}) t} ~ |\mu_j|^2}  \right)},
      \label{eq:mle_closed_form}
\end{equation}
where $\lVert \bm{x}(0)\rVert$ is the Euclidean norm of the initial state $\bm{x}(0)$, $\gamma_{j}$ is the $j$-th eigenvalue of $\bm{K}$, and $\mu_{j} = \langle \bm{x}(0), \bm{v}_{j} \rangle$  is the projection of the initial state onto the eigenvectors of $\bm{K}$, with $\langle . \rangle$ representing the complex-valued inner product. Similarly, $\nu_j$ is the projection of a state representing the initial deviation between nearby trajectories onto the eigenvectors of $\bm{K}$. 

Equation (\ref{eq:mle_closed_form}) provides a closed-form expression for the time-dependent MLE as a function of the initial state of the system $\bm{x}(0)$ and the eigensystem of the matrix $\bm{K}$, which contains information about the network connectivity and delayed interactions. This equation, in turn, allows calculating the time-dependent MLE for individual dynamical trajectories in the system. The theoretical and numerical calculations both agree and, further, demonstrate that $\lambda$ exhibits positive values during the transient (Fig.~\ref{fig:schematic_examples}f; Supplement Fig.~S2). As the network transitions to the fixed point, $\lambda$ becomes negative. These negative values of $\lambda$ are consistent with the values expected for a system in the vicinity of a fixed point. The positive values during the period of rich spatiotemporal dynamics, however, are consistent with transient chaos. These results provide the first closed-form theoretical expression for the MLE in a large nonlinear network, allowing us to demonstrate that rich spatiotemporal activity patterns with positive MLE emerge during the regimes where we have previously observed that the network can perform challenging computations \cite{budzinski2024exact}.

\begin{figure*}[t]
    \centering
    \includegraphics[width=\linewidth]{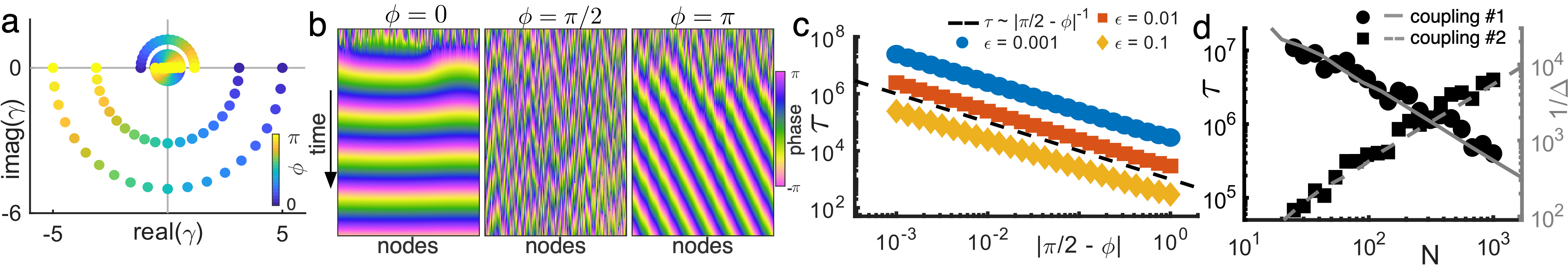}
    \caption{\textbf{The role of phase-delay and network size on transient lifetimes.} \textbf{(a)} The phase-delay $\phi$ leads to a rotation of the eigenvalues of $\bm{K}$. \textbf{(b)} The phase dynamics of the network displays a phase synchronized state for $0 \leq \phi < \sfrac{\pi}{2}$; remains asynchronous for $\phi = \sfrac{\pi}{2}$; and displays a wave pattern for $\sfrac{\pi}{2} < \phi \leq \pi$. \textbf{(c)} We observe a power-law decay of the transient lifetime $\tau$ as $|\sfrac{\pi}{2} - \phi|$ increases (average over $100$ realizations for both $\phi > \sfrac{\pi}{2}$ and $\phi < \sfrac{\pi}{2}$) for different coupling strength values $\epsilon$. \textbf{(d)} We fix the coupling strength ($\epsilon_1 = 0.001$ and $\epsilon_2 = 100$) and the delay parameter $\phi = \sfrac{\pi}{2} -0.01$ and vary $N$ in two different conditions: ``coupling \#1" with $\bm{K} = \epsilon_1 e^{-\i \phi} \bm{A}$, where the transient lifetime $\tau$ decreases with $N$ (black circles) and ``coupling \#2" with $\bm{K} = (\sfrac{\epsilon_2}{N^2}) e^{-\i \phi} \bm{A}$, where $\tau$ increases with $N$ (black squares). Here, we consider $k$-ring graphs with $k = \mathrm{floor}(0.25N)$ and plot the average over $100$ realizations. The inverse of real part of the spectral gap $\Delta = \mathrm{Re}(\gamma_1 - \gamma_2)$ of $\bm{K}$ is a predictive of the transient lifetime in both cases (solid and dashed lines).}
    \label{fig:lifetime}
\end{figure*}
With these results in mind, one can now ask an important question: are these transients long-lived enough to be useful in real-world computations? We can understand this theoretically in terms of the spectrum of the network: specifically, the connectivity $\bm{A}$ and the phase-delay $\phi$ work together to determine specific configurations of the eigenvalues of the aggregate matrix $\bm{K}$. We can study this fully analytically in the case of networks defined on $k$-ring graphs, where the Circulant Diagonalization Theorem provides explicit formulae for the eigenvectors and eigenvalues \cite{davis1979} (see Supplement Sec.~V). $\phi$ then determines a one-parameter rotation of the eigenvalues in the complex plane (Fig.~\ref{fig:lifetime}a). By varying $\phi$, then, we can control the relationships between the eigenvalues of $\bm{K}$ and, hence, the network dynamics: specifically, for $0 \leq \phi < \sfrac{\pi}{2}$ the network will synchronize in phase before reaching the asymptotic state; for $\phi = \sfrac{\pi}{2}$, the network will remain asynchronous; and for $\sfrac{\pi}{2} < \phi \leq \pi$ the network will display a wave pattern in its phase dynamics (Fig.\ref{fig:lifetime}b). $\phi$ is, then, a  bifurcation parameter that controls the asymptotic state and how trajectories in the network flow to that state over the course of the transient. We note that this result is general to any Hermitian $\bm{A}$.

The parameter $\phi = \sfrac{\pi}{2}$ then becomes a critical bifurcation point that leads to different behavior. At this point, the eigenvalues of $\bm{K}$ become purely imaginary. This results in unique network dynamics where the amplitude remains finite, the collective dynamics is fully asynchronous, and the network never reaches the fixed point at the origin. Further, the closer that $\phi$ comes to this critical value, the longer the transient lifetimes become. This behavior follows a power-law relationship with the distance from $\sfrac{\pi}{2}$, for both cases $\phi > \sfrac{\pi}{2}$ and $\phi < \sfrac{\pi}{2}$ (Fig.~\ref{fig:lifetime}c). Studying this behavior in $\phi$ thus provides theoretical and mechanistic insight into controlling long transients in the network that can, in turn, be useful for computation. 

We next studied how the transient lifetime varies with network size. For a fixed coupling strength $\epsilon$ and phase delay $\phi$ in the $k$-ring graph, the transient lifetime $\tau$ decreases with increasing $N$ (black circles, Fig.~\ref{fig:lifetime}d). This scaling behavior can be clearly understood through the spectral properties of the matrix $\bm{K}$, because the spectral gap $\Delta = \mathrm{Re}(\gamma_1 - \gamma_2)$ increases with $N$, driving the mode associated with synchronization ($\bm{v}_1$) to dominate more rapidly. The inverse of the spectral gap $\sfrac{1}{\Delta}$ accurately predicts the transient length $\tau$, providing mechanistic insight into how $\tau$ varies with $N$ (solid gray line, Fig.~\ref{fig:lifetime}d). In addition, however, this analytical insight allows us to obtain networks that now show the opposite behavior: by choosing a coupling scheme that is inversely proportional to the $N^2$, we can obtain a system where the transient lifetime increases with $N$ (black squares, Fig.~\ref{fig:lifetime}d), which is again explained by the spectral gap $\Delta$ (dashed gray line, Fig.~\ref{fig:lifetime}d). Taken together, these results show that our mathematical framework provides a precise link between network parameters and dynamics through the spectrum of $\bm{K}$, which opens the possibility of designing transient dynamics with different scaling properties.

\begin{figure}[b]
    \centering
    \includegraphics[width=0.875\linewidth]{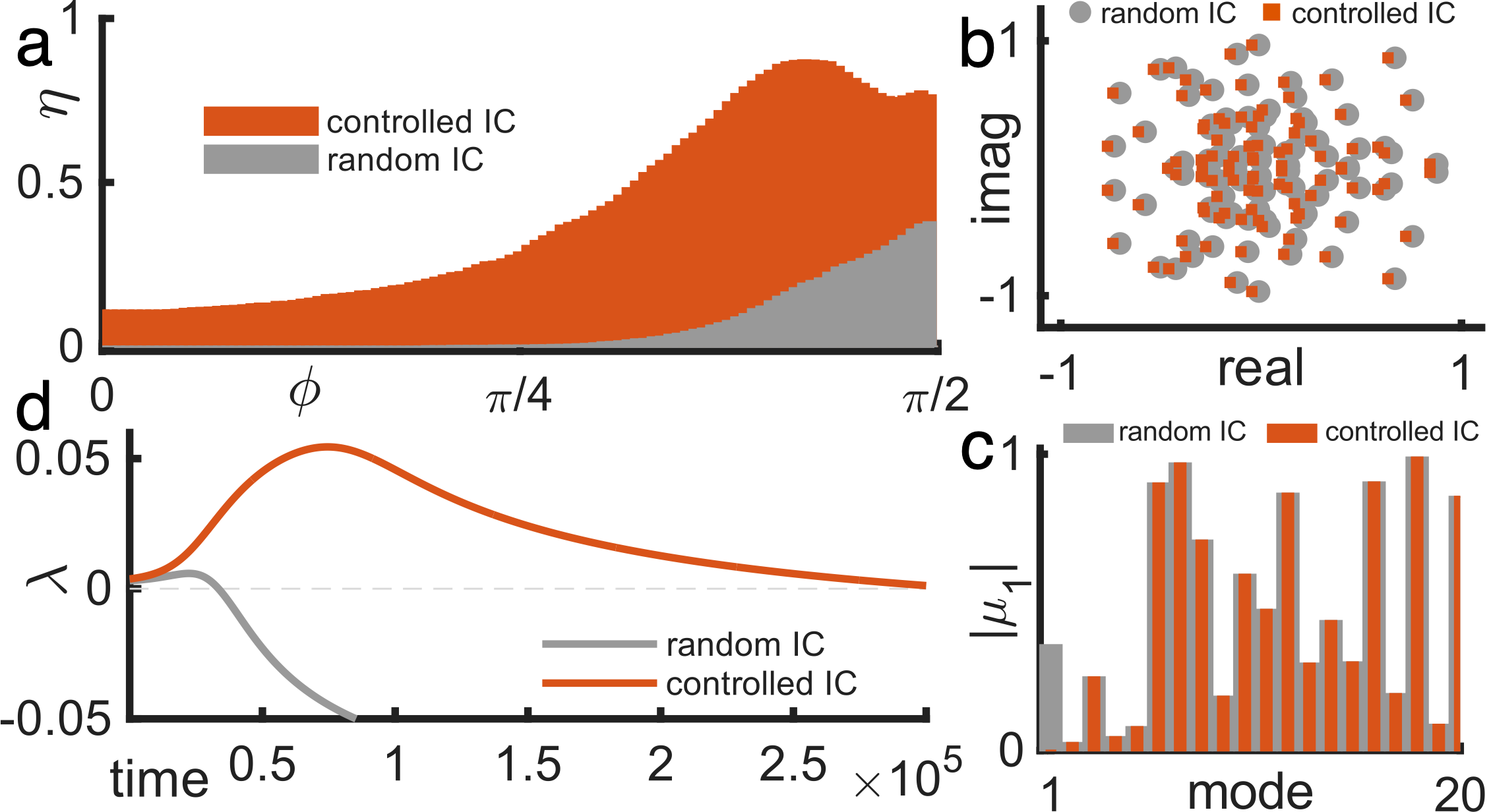}
    \caption{\textbf{Initial conditions and transient dynamics.} \textbf{(a)} We initialize the network on a random state and evaluated the fraction of realizations $\eta$ (over $10,000$ realizations) that lead to $\lambda > 0$ at $t = 10^4$ for different values of $\phi$ (gray bars). We also consider a set of ``controlled initial states", with the condition $|\mu_1| = 0$ (no contribution of the synchrony mode). This leads to a much bigger ratio of realizations with $\lambda > 0$ at $t = 10^4$ (orange bars). \textbf{(b)} The controlled initial states are very similar to the fully random ones (small shift in the real axis), \textbf{(c)} with only difference being the projection onto the synchrony eigenvector $|\mu_1|$. \textbf{(d)} These almost identical initial states lead to very different behavior during transient. Here, we consider $k$-ring graphs with $N = 101$ and $k = 25$, $\omega = 0$, and $\epsilon = 0.025$.}
    \label{fig:initial_condition_mle}
\end{figure}
We note, finally, that the projection of the initial condition $\bm{x}(0)$ onto the first eigenvector $\bm{v}_1$ of $\bm{K}$, $\mu_{1} = \langle \bm{x}(0), \bm{v}_1 \rangle$, critically determines the lifetime of the transient. This is apparent from a strong negative correlation between the value of this projection and the lifetime $\tau$, in addition to the MLE (Supplement Fig.~S3). Leveraging this analytical insight, we can then design a strategy to control the transient lifetime through an algebraic manipulation of the initial condition $\bm{x}(0)$. We first project an initial condition into the eigenbasis of $\bm{K}$, obtaining $\bm{\hat{x}}(0) = V^\dagger\bm{x}(0)$. We next set $(\bm{\hat{x}}(0))_1 = 0$, removing the contribution from the eigenvector associated with phase synchronization $\bm{v}_1$ to the initial state. We then project the initial condition back into the original basis. This algebraic procedure results in a modified initial condition vector $\bm{x}^{\prime}(0)$ that has no projection onto $\bm{v}_1$. When tested on 10,000 independent initial conditions, this algebraic procedure dramatically increases the fraction of realizations producing transients with positive time-dependent MLE (orange bars, Fig.~\ref{fig:initial_condition_mle}a) compared to unconstrained random states (blue bars). Remarkably, the controlled initial conditions appear indistinguishable from random states in the complex plane, with only a slightly difference in the real axis, which results from the projection onto $\bm{v}_1$ (Figs.~\ref{fig:initial_condition_mle}b \ref{fig:initial_condition_mle}c). Despite this subtle difference, they produce vastly different network behavior: the algebraically modified states generate long transients with positive MLE, while the original states can yield short transients (Fig.~\ref{fig:initial_condition_mle}d). These results demonstrate that our analytical framework enables precise control over both network parameters and initial conditions to reliably generate long-lived transients with rich spatiotemporal dynamics. 

Standard mathematical approaches have made substantial progress in describing the asymptotic, or long-term, behavior of complex systems. This approach is highly effective in physics because systems like a gas in a flask settle to equilibrium so rapidly that transient states are rarely observed. Asymptotic methods have proven invaluable to analyze a range of these systems, including large networks of nonlinear oscillators. As a specific example, the Master Stability Function has revolutionized our understanding of the stability of the synchronous state in networked systems \cite{pecora1998master,barahona2002synchronization,belykh2004connection,hart2019topological,sugitani2021synchronizing}.

When considering computation in neural networks, however -- either biological or artificial -- these systems are continuously receiving inputs and computing in dynamic, transient, and non-equilibrium states \cite{maass2002real,rabinovich2008transient,mazor2005transient}. The physics of transient states has been a focus of interest for many years \cite{grebogi1983fractal,grebogi1983crises,kantz1985repellers,grebogi1986critical,tel2008chaotic,lai2011transient}. The possibility for rich dynamics during transients, including transient chaos, has been observed across many systems \cite{tel2008chaotic, tel2015joy, wacker1995transient} and can be explained in terms of bifurcations and topological properties of the phase space \cite{grebogi1983crises, grebogi1983fractal,lai2011transient,koch2024ghost}, and the possibility of controlling chaotic dynamics has been a major focus of research \cite{ott1990controlling,pyragas1992continuous,boccaletti2000control,liu2016control}. At the same time, however, an exact mathematical description of transient dynamics during computation has been lacking.

This work provides an advance in the analytical understanding of computation with transient states. We obtain an exact equation for the time-dependent MLE and demonstrate that this system exhibits positive Lyapunov exponents during the transient periods that are also useful for computation. This is linked to the rich diversity of spatiotemporal dynamics the network can produce, including chimera-like states with synchronous and incoherent clusters in the phase dynamics. Chimera states have previously been shown to be chaotic transients in networks of Kuramoto oscillators at the finite scale \cite{wolfrum2011chimera,omel2010chimera}, and our results are reminiscent of these previous works, while also offering an exact mathematical analysis that is valid both for individual networks and individual dynamical trajectories.

This framework allows, for the first time, analytically determining the network conditions -- including connectivity, coupling parameters, and specific initial conditions -- that produce transients with positive MLE, and with specific lifetimes. By providing precise mathematical tools to characterize the dynamical regimes involved in transient computation, this framework opens new opportunities for understanding how these neural networks compute. While recent theoretical work has made significant progress in analyzing the learning dynamics of deep neural networks \cite{saxe2013exact}, including deriving Lyapunov exponents for weight evolution during training of transformers \cite{cowsik2025geometric}, these studies focus primarily on the training process and the features that enable efficient learning. The mathematical approach we introduce here is complementary: rather than analyzing how networks learn, we provide tools to analyze the forward pass of trained networks -- the individual computations where a specific input drives the network to generate a specific output. Our work thus offers a new lens for studying computation in these neural networks by connecting their complex input-output mappings to a mathematical framework of transient chaos.

\begin{acknowledgements}
This work was supported by BrainsCAN at Western University through the Canada First Research Excellence Fund (CFREF), the NSF through a NeuroNex award (\#2015276), the Natural Sciences and Engineering Research Council of Canada (NSERC) grant R0370A01, Compute Ontario (computeontario.ca), Digital Research Alliance of Canada (alliancecan.ca), and NIH Grants U01-NS131914 and R01-EY028723. R.C.B acknowledges the support of the Canada Research Chairs program.
\end{acknowledgements}

\end{document}